\documentclass[aps,10pt, pre,twocolumn,superscriptaddress]{revtex4-1}
\usepackage{graphicx}
\usepackage{float}
\usepackage{amsmath}
\usepackage{mathtools} 
\usepackage{color}
\usepackage{afterpage}
\usepackage{xr}
\usepackage{makecell}

\usepackage{amssymb}
\usepackage{graphicx,color} 
\usepackage{bbm} 
\usepackage{multirow}
\usepackage{enumitem}
\usepackage{mathrsfs}
\usepackage{commath}
\usepackage{hyperref}
\usepackage{fontenc}
\usepackage[dvipsnames]{xcolor}

\usepackage{tabularx}

\usepackage[english]{babel}
\usepackage{float}
\usepackage{amsmath,amssymb}
\usepackage{mathtools} 
\usepackage{color}
\usepackage{xr}
\usepackage{graphicx}
\usepackage{dcolumn}
\usepackage{bm}

\newcolumntype{Y}{>{\centering\arraybackslash}X}

\begin{document}

\preprint{APS/123-QED}

\title{The dimensional evolution of structure and dynamics in hard sphere liquids}

\author{Patrick Charbonneau}
\affiliation{Department of Chemistry, Duke University, Durham, North Carolina 27708}
\affiliation{Department of Physics, Duke University, Durham, North Carolina 27708}
\author{Yi Hu}
\affiliation{Department of Chemistry, Duke University, Durham, North Carolina 27708}
\author{Joyjit Kundu}
\affiliation{Department of Chemistry, Duke University, Durham, North Carolina 27708}
\affiliation{Department of Physics, Indian Institute of Technology Hyderabad, Kandi, Sangareddy, TN 502285, India}
\author{Peter K. Morse}
\thanks{Corresponding author}
\email{peter.k.morse@gmail.com}
\affiliation{Department of Chemistry, Duke University, Durham, North Carolina 27708}

\date{\today}

\begin{abstract}
The formulation of the mean-field, infinite-dimensional solution of hard sphere glasses is a significant milestone for theoretical physics. How relevant this description might be for understanding low-dimensional glass-forming liquids, however, remains unclear. These liquids indeed exhibit a complex interplay between structure and dynamics, and the importance of this interplay might only slowly diminish as dimension $d$ increases. A careful numerical assessment of the matter has long been hindered by the exponential increase of computational costs with $d$. By revisiting a once common simulation technique involving the use of periodic boundary conditions modeled on $D_d$ lattices, we here partly sidestep this difficulty, thus allowing the study of hard sphere liquids up to $d=13$. Parallel efforts by Mangeat and Zamponi [Phys.~Rev.~E \textbf{93}, 012609 (2016)] have expanded the mean-field description of glasses to finite $d$ by leveraging standard liquid-state theory, and thus help bridge the gap from the other direction. The relatively smooth evolution of both structure and dynamics across the $d$ gap allows us to relate the two approaches, and to identify some of the missing features that a finite-$d$ theory of glasses might hope to include to achieve near quantitative agreement.
\end{abstract}

\maketitle

\section*{Introduction}

Hard spheres are a classical minimal model for the structure and dynamics of real (yet simple) liquids~\cite{hansen_theory_2013}. The model captures many of the salient features of liquids, while being simple enough to treat using both analytical theory and computer simulations. From a theoretical standpoint, in particular, there have been two main thrusts: (i) developing models that capture the features of low-dimensional ($d=2$ and $d=3$) fluids reasonably well, and (ii) developing a finite-$d$ mean-field description that becomes exact in the limit $d\rightarrow\infty$~\cite{parisi_mean-field_2010,parisi_theory_2020}. Both approaches have been met with success~\cite{hansen_theory_2013,berthier_theoretical_2011,charbonneau_glass_2017, parisi_theory_2020},
but the extrapolation of the former to high densities and of either to intermediate dimensions (e.g., ${4\le d < 20}$) leads to various quantitative inconsistencies. 

In order to expand on this point, consider first the high-density regime. A low volume fraction ($\varphi$) Brownian fluid of hard spheres begins as a Fickian fluid characterized by a purely diffusive mean squared displacement (MSD). Upon compression, non-Fickian diffusion first emerges at a dynamical onset, $\varphi_\mathrm{nf}$~\cite{saltzman_activated_2006, kurchan_exact_2013, charbonneau_hopping_2014, manacorda_numerical_2020}. Upon further compression, liquid dynamics turns increasingly sluggish as particles first become transiently caged by each other, and then nearly arrested. In mean-field descriptions, this sustained caging is associated with a topological change in the free energy landscape at the dynamical (or mode-coupling theory) transition, $\varphi_\mathrm{d}$~\cite{berthier_theoretical_2011, charbonneau_glass_2017}. Although this transition is avoided in finite-$d$ systems, the underlying crossover has long been extracted from the pseudo-divergence of the structural correlation time observed in high density fluids~\cite{gotze_complex_2008,berthier_theoretical_2011,coslovich_localization_2019}.

Signatures of both $\varphi_\mathrm{d}$ and $\varphi_\mathrm{nf}$ have long been sought out in the intricate structure of low-dimensional liquids, but $d\rightarrow\infty$ fluids contain no such feature. For hard spheres, the pair correlation function is then flat beyond contact, and all higher-order correlations are trivially factorizable~\cite{parisi_theory_2020}. The structure of low-dimensional liquids, which is expected to smoothly evolve towards the high-dimensional description, must therefore at least quantitatively perturb the mean-field predictions for these two signature features. 

The chasm between what has been extracted from existing numerical simulations and the $d\rightarrow\infty$ solution is, however, too vast to say for sure whether this proposal holds. For instance, theoretical estimates for the dynamical transition computed using both the Percus-Yevick (PY) and Hypernetted Chains (HNC) approximations in $d=2-70$~\cite{mangeat_quantitative_2016} deviate substantially from each other and from simulation estimates~\cite{charbonneau_glass_2011, charbonneau_dimensional_2012}. While both HNC and PY capture the $d\rightarrow\infty$ value of $\varphi_\mathrm{d}$, there is no indication that this high-dimensional approach is the same as that of HS or thus that the gap between simulation and these two theories tends vanish with increasing $d$.
The observed discrepancy might partly result from the small system sizes used in simulations, but might also attest to the structural inadequacies of the liquid-state descriptions. Given that our understanding of glasses and supercooled liquids has since developed, can the dimensional gap be properly bridged? Hardware improvements over the last decade offer some hope on the computational side, but remain far from sufficient. Fortunately, algorithmic techniques have also since been refined. In particular, the development of smart boundary conditions for high-$d$ systems have significantly extended the numerical reach~\cite{charbonneau_thermodynamic_2021}. 

These advances here allow us to reach near quantitative agreement between theory and numerics in some respects. More specifically, in this work we show that the relevant features of hard sphere liquids converge to those of HNC fluids as $d$ increases, and that for $d \geq 13$ they become almost indistinguishable, thus possibly making the study of higher-dimensional fluids essentially unnecessary. We also show that the HNC-based predictions for $\varphi_\mathrm{d}$ tend to track the numerical results, but remain quantitatively distinct. The rest of this paper is structured as follows. Section~\ref{sec:methods} details the numerical techniques that allow us to equilibrate dense liquids up to $d=13$. Section~\ref{sec:dynamics} describes the evolution of fluid pair structure with dimension, Section~\ref{sec:dyn} describes the evolution of dynamics with dimension, and Section~\ref{sec:conclusion} concludes by briefly considering how further theoretical advances might strengthen the quantitative interplay between the mean-field theory of glasses and numerical simulations. 

\section{Model and Methods}
\label{sec:methods}
Throughout this work, we consider hard sphere liquids composed of $d$-dimensional hyperspheres of diameter $\sigma$, with $4\leq d \leq 13$. For $d \ge 4$, monodisperse spheres suffice because crystallization, while thermodynamically possible~\cite{charbonneau_thermodynamic_2021, charbonneau_three_2021}, is then effectively always suppressed~\cite{skoge_packing_2006, van_meel_hard-sphere_2009,van_meel_geometrical_2009, charbonneau_numerical_2010}. Initial states are first prepared using Poisson-distributed soft hyperspheres whose energy is minimized to zero, thus yielding a valid hard sphere configuration. These configurations are then equilibrated via standard Monte Carlo (MC) dynamics~\cite{frenkel_understanding_2001}, using attempted individual particle step sizes of $\kappa \delta \sigma$ along each dimension with $\kappa\in[-1,1)$ distributed uniformly at random and $\delta \approx 1/(6d)$ chosen so as to optimize the equilibration time at high $\varphi$. The numerical results are however  rather insensitive to this specific choice, given the rescaling described in Sec.~\ref{sec:phid}. In the rest of this section, we describe various aspects of the numerical implementation, paying particular attention to features that enable the efficient consideration of higher-dimensional systems, such as the simulation box shape and the system size $N$.

\subsection{Equilibration and Sampling}
\label{sec:dynamics}
The timescale for structural decorrelation (and hence equilibration) is determined from the characteristic decay of the particle-scale self-overlap
\begin{equation}
Q(t) = \frac{1}{N} \sum^N_{i=1} \Theta(a\sigma - |\mathbf{r}_i(t) - \mathbf{r}_i(0)|)
\label{eqn:overlapFunction}
\end{equation}
where $r_i(t)$ is the position of particle $i$ at time $t$, $\Theta$ is the Heaviside function, and $a$ is a length on the scale of the typical particle cage size around $\varphi_\mathrm{d}$. (In this work, $a = 0.3$. Although asymptotic $1/d$ corrections to $a$ are expected, in finite-$d$ the dependence has been found to be relatively weak~\cite{berthier_finite_2020}. In any event, taking a constant value merely overestimates $\tau_\alpha$, which has no bearing on the subsequent analysis, as further discussed in Sec.~\ref{sec:phid}.) In other words, $Q(t)$ captures the fraction of particles having moved a distance larger than $a\sigma$ at time $t$. This function is known to generically decay as a stretched exponential ${Q(t) \sim \exp(-(t/\tau_\alpha)^\zeta)}$ with $\zeta=1$ for Fickian fluids and $\zeta$ decreasing as $\varphi_\mathrm{d}$ is approached. The characteristic relaxation time $\tau_\alpha$ is then implicitly defined as $Q(\tau_\alpha)=1/e$. 

Lower density systems are equilibrated for at least $3\tau_\alpha$, before recording the pair correlation function, $g(r)$, and the mean squared displacement (MSD), $\langle r^2(t) \rangle$. This starting configuration is also used for generating additional equilibrated configurations, whereupon $1000$ realizations are averaged. To properly characterize higher density configurations, and in particular to extract the dynamical transition, simulations need to be run for much longer as $\varphi_ \mathrm{d}$ is approached. An equilibration of at least $30\tau_\alpha$, and as few as $10$ independent realizations are then used.

\subsection{Structure and Pressure}
\label{sec:struct}

For distances $r$ less than half the box size, the standard scheme to extract $g(r)$ is used (see, e.g., Ref.~\cite[Sec.~4.4]{frenkel_understanding_2001}), because the shell of neighbors is then perfectly spherical. For $r$ larger than half the box size the shell encompassing $r$ intersects the periodic box limits. Although the particle count and the shell volume are then both smaller than in perfect spherical shell, the sphere density is unaffected and thus relevant structural information can still be extracted. Because of the relatively complex geometry of the truncated shell, however, its volume is obtained by simple Monte Carlo integration rather than analytically. 

From the virial theorem, the reduced pressure $p$ for hard spheres can be determined from the contact value of the pair correlation,
\begin{equation}
\label{eq:eos}
p(\varphi) = \frac{\beta P}{\rho} = 1 + 2^{d-1}\varphi g(\sigma^+)
\end{equation}
where $\beta = 1/k_BT$ is the inverse temperature , $P$ is the standard pressure, and $\rho = N/V$ is the number density for a simulation box of volume $V$. In order to compare thermodynamic quantities across dimensions, we correct for their asymptotic ${d \to \infty}$ scaling. In particular, the volume fraction is rescaled as $\widehat{\varphi} = 2^d\varphi/d$ and the reduced pressure as $\widehat{p} = p/d$. Equation~\eqref{eq:eos} then simplifies to
\begin{equation}
\widehat{p}(\widehat{\varphi}) = \frac{1}{d} + \frac{\widehat{\varphi}}{2} g(\sigma^+).
\label{eq:eos}
\end{equation}
This rescaling makes clear that, given that $g(r)$ is but a pure step function in limit $d\rightarrow\infty$, the equation of state then becomes~\cite{wyler_hard-sphere_1987} (see also Ref.~\cite[Eq.~2.61]{parisi_theory_2020})
\begin{equation}
\label{eqn:eosdinfty}
\widehat{p}=\frac{\widehat{\varphi}}{2} \quad \textrm{for} \quad d\to \infty.
\end{equation}
%

\begin{figure}[h]
\includegraphics[width=0.9\linewidth]{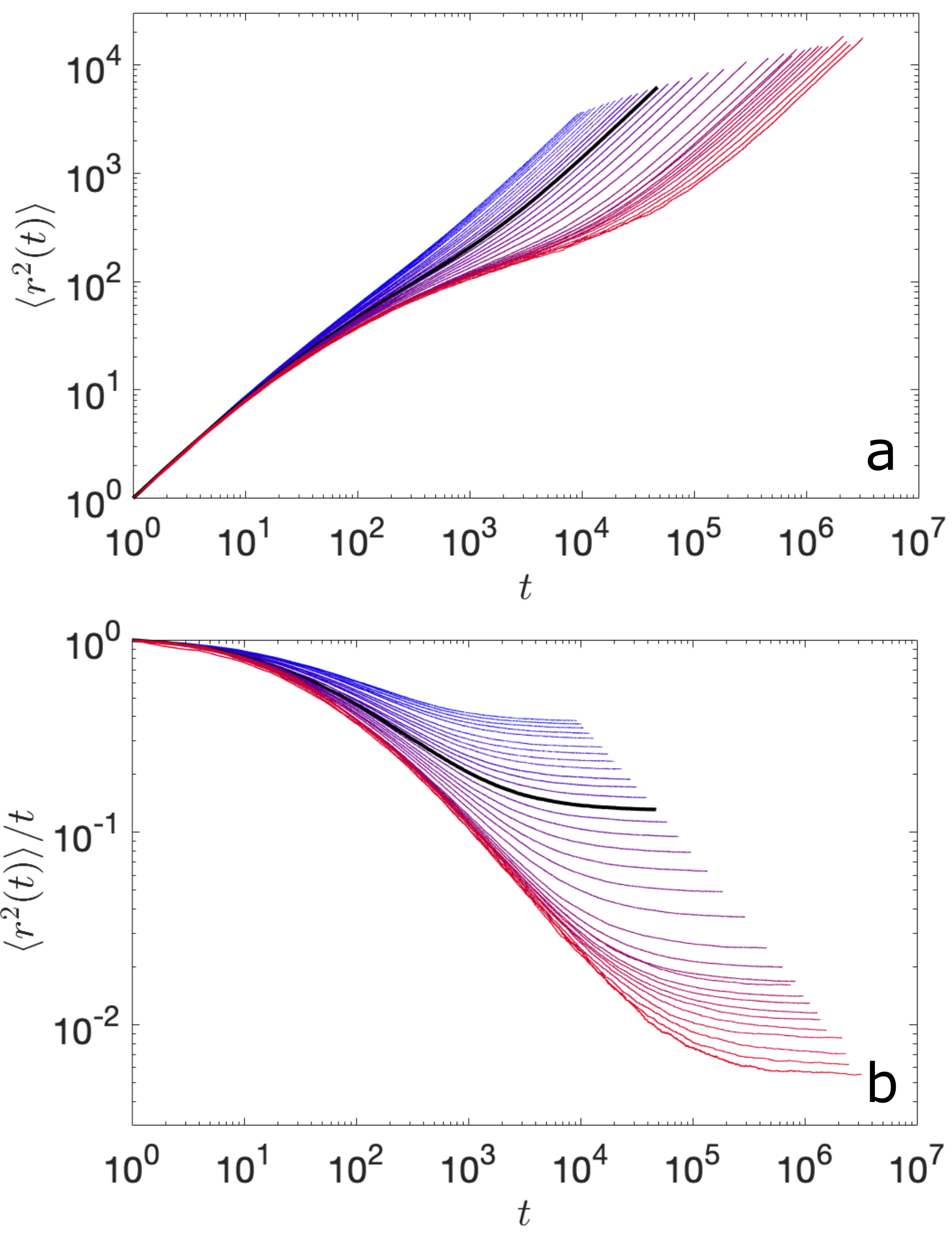}
\caption{\textbf{a)} MSD of a hard sphere fluid in $d=5$ averaged over 1000 realizations. The system smoothly goes from Fickian at $\widehat{\varphi}=1$ (blue), through the onset of non-Fickian diffusion at $\varphi_\mathrm{nf}=1.36$ (black line), and up to a dense fluid $\widehat{\varphi}=1.648$ (red) near the dynamical transition (here $\widehat{\varphi}_\mathrm{d} = 1.706$). \textbf{b)} The diffusion constant $D$ can be extracted from the long-time plateau of $\langle r(t) \rangle/t$ after rescaling the short-time dynamics. Fits to a standard exponential decay form are overlaid.}
\label{fig:phiDphiOnset}
\end{figure}

\subsection{Boundary Conditions}
As mentioned in Sec.~\ref{sec:struct}, the standard scheme for determining the radial distribution function is valid up to the inscribed sphere radius of the periodic box. Beyond this point, a particle may have multiple periodic images of a same neighbor within its spherical shells, which gives rise to unphysical structural correlations. In this sense, the choice of periodic boundary conditions is optimized when the inscribed sphere radius is maximal for a given box volume, under the periodic tiling constraint of $\mathbb{R}$. Optimizing boundary conditions is thus equivalent to optimizing sphere packing in a given $d$. The conventional cubic periodic boundary condition, which corresponds to a simple cubic tiling, is clearly not optimal, because this sphere packing lattice is not the densest in any $d \ge 2$.

This realization is far from novel. The efficiency of non-cubic periodic boxes attracted interest historically when computer resources were limited~\cite{adams_computer_1979,allen_computer_1989}.
Yet the resulting efficiency improvement was not found to be sufficiently significant in $d=3$ ($\approx40\%$), in the context of rapidly accelerating computer hardware, especially given the coding apprehension such approach involves~\cite{frenkel_simulations_2013}. 
As $d$ increases, however, hypercubic periodic boxes become increasingly inefficient, and non-cubic lattice packings can give rise to more sizable efficiency gains. Realizing that the key implementation for simulating some of these boxes can be straightforwardly adapted from quantizing algorithms in information theory~\cite{conway_fast_1982} further motivates their reconsideration. For a common simulation code structure~\cite{frenkel_understanding_2001}, the only significant algorithmic change concerns the minimal image convention, and the structure of that change is often independent of $d$.

More specifically, we first construct the periodic box scheme for $Z_d$ (cubic) and $D_d$ (checkerboard, such as face-centered cubic in $d=3$) based lattices. We denote $f(x)$ the closest integer to $x$, and likewise $f(\bm{x}) = (f(x_1), ..., f(x_n))$ for an $n$-dimensional vector $\bm{x}$. Clearly, $\delta{\bm{x}} = \bm{x} - f(\bm{x})$ is the minimum image of $\bm{x}$ to the origin, in (hyper-)cubic periodic boundary conditions, $\mathbb{Z}^n$, and computing it requires $d$ operations. We also define $f_2(x)$ as the second-nearest integer to $x$, and 
\begin{equation}
\begin{aligned}
f_2(\bm{x}) = (f(x_1), f(x_2), ..., f_2(x_k), f(x_{k+1}), ..., f(x_d) ),\\
 \quad \lvert  x_k - f(x_k)\rvert\ \ge \lvert x_i-f(x_i) \rvert\ \forall i
\end{aligned}
\end{equation}
is the second-nearest integer vector to $\bm{x}$ in $\mathbb{Z}^d$. The $D_d$ minimum image of $\bm{x}$ is chosen to be
\begin{equation}
\delta(\bm{x}) = \begin{cases}
\bm{x} - f(\bm{x}), & \sum_{i=1}^d f(x_i) \text{ is even,} \\
\bm{x} - f_2(\bm{x}), & \sum_{i=1}^d f(x_i) \text{ is odd.}
\end{cases}
\end{equation}
This construction defines the $D_d$ periodic box, and its calculation requires roughly $4 d$ operations~\cite{conway_soft_1986}. Additionally, the $D_d$ lattice has a maximum inscribed radius ${R_I = 1/\sqrt{2}}$ and a circumradius of either 1 for $d \le 4$ or $\sqrt{d}/2$ for $d > 4$. Therefore, the volume of a $D_d$ periodic box is 2~\cite{conway_sphere_1993}, and the packing fraction of the $D_d$ lattice is $2^{\frac{d-2}{2}}$ times greater than $\mathbb{Z}^d$, yielding an efficiency gain of that same factor. 

The furthest points to lattice sites in $D_d$ lattices (known as deep holes) are at $(\frac{1}{2}, ..., \frac{1}{2})$ and its periodic images.
The distance of the deep hole to a lattice site, $\sqrt{d}/2$, becomes no less than twice of the inscribed radius for $d \ge 8$. Therefore, one can then slide a second $D_d$ lattice within these holes without changing the inscribed radius of the Voronoi tessellation. The volume of a periodic box is then cut in half. This construction results in $D_d^+$ lattices, whose symmetry is only valid in even dimensions. The minimum image of $\bm{x}$ in $D_d^+$ lattice is either $\delta{\bm{x}}$ or $\delta(\bm{x}-\frac{\bm{1}}{\bm{2}}) + \frac{\bm{1}}{\bm{2}}$, whichever has the smaller norm. 
In particular, taking $d=8$ gives an $E_8$ periodic box and also corresponds to the densest sphere packing in that dimension. 
For odd dimensions in $d \ge 9$, one can also slide another copy of $D_d$, but now by an offset of $\bm{a} = (\frac{1}{2}, ..., \frac{1}{2}, a_d )$, where $a_d \in \mathbb{R}$ is an arbitrary real number.
By convention, taking $a_d=0$ gives the lattice $D_d^{0+}$. The minimum image of $\bm{x}$ in $D_d^{0+}$ lattices is either $\delta{\bm{x}}$ or $\delta(\bm{x}-\bm{a}) + \bm{a}$, whichever has the smaller norm. 
In particular, $D_9^{0+}$ is a realization of the $\Lambda_9$ lamellar lattice, which is the densest known sphere packing structure in $d=9$.

\begin{table}[h]
\caption{Summary of select periodic box properties}
\begin{tabularx}{\linewidth}{YYYYY}
\hline \hline
Symmetry & Valid in & Volume ($V$) & \thead{Inscribed \\ radius \\ ($R_I$)} & \thead{Packing \\ efficiency \\ w.r.t. $\mathbb{Z}^d$}  \\
\hline
$\mathbb{Z}_d$ & $d\geq 1$ & 1 & 1/2 & 1\\
$D_d$ & $d \ge 2$ & 2 & $1/\sqrt{2}$ & $2^{\frac{d-2}{2}}$  \\
 $D_d^+$ & even $d \ge 8$ & 1 & $1/\sqrt{2}$ & $2^{\frac{d}{2}}$  \\
  $D_d^{0+}$ & odd $d \ge 9$ & 1 & $1/\sqrt{2}$ & $2^{\frac{d}{2}}$  \\
  $\Lambda_{24}$ & $d=24$ & 1 & 1 & $ 2^{24}$ \\
  \hline
\end{tabularx}
 \label{tab:pbc}
\end{table}

The inscribed radius of both $D_d^+$ and $D_d^{0+}$ periodic boxes remains $1/\sqrt{2}$, and their volume is 1---half that of $D_d$ boxes.
The implementation of $D_d^+$ and $D_d^{0+}$ periodic box thus doubles the efficiency over $D_d$. Because it involves computing two sets of $D_d$ minimum images and squared distances, the minimum image algorithm takes $13 d$ operations in total. (For the $E_8$ lattice a special quantization algorithm takes only 72 operations~\cite{conway_soft_1986}.) 
In practice, we observe that the minimum image determination is roughly three times as computationally demanding as for $D_d$.
The efficiency gain of these lattices is thus canceled if the minimum image determination is the computational bottleneck, as is the case in liquid simulations. (It may be possible to do better in certain higher dimensions, such as for the remarkably dense Leech lattice, $\Lambda_{24}$, which is a factor of $2^{13}$ denser than $D_{24}$, and for which a fast quantization algorithm takes \emph{only} 55,968 steps~\cite{conway_soft_1986}. A factor of roughly 10---compared to $4d \times 2^{13}=786,432$---should thus be gained from this geometry.) In practice, we thus use simulation boxes with standard cubic periodic boundaries in $d=3$ and $d=4$, and $D_d$ periodic boxes for $d \ge 5$.

\subsection{System Size Considerations}
A particularly subtle issue in higher dimensional simulations is that of system size. Careful consideration has thus been given to this matter in  prior simulation studies~\cite{charbonneau_glass_2011,charbonneau_geometrical_2013}. For the sake of the current work, it suffices to recall that for proper thermodynamic quantities to be extracted, systems should be much larger than the largest correlation length. Once that target is attained, remaining corrections should scale as $1/N$. Hence, as long as the systems simulated are larger than the static and dynamical correlation lengths (hydrodynamic correlations are not relevant in MC simulations because momentum is not conserved), then only trivial system size corrections should persist. 

In low-dimensional dense fluids, various static features have been argued to play a key dynamical role. The slow convergence of the virial series further suggests that multi-particle correlations also then contribute significantly to the equation of state. As $d$ increases, however, structural correlations are expected to steadily vanish and structural anomalies that couple to dynamics are expected to homogenize, at least for $\widehat{\varphi}<\widehat{\varphi}_\mathrm{d}$. Although the dynamical correlation length is expected to diverge at $\widehat{\varphi}_\mathrm{d}$ -- as the crossover hardens with increasing $d$ -- the critical regime is expected to be fairly narrow~\cite{charbonneau_glass_2011, charbonneau_hopping_2014}. 

Therefore, as long as we avoid introducing correlations by, say, having a particle interact with itself through the periodic boundary, then relatively modest simulation box sizes should suffice. For these purposes, it is useful to note that the shortest distance across the simulation box is $2R_I$, where $R_I$ is the maximum inscribed radius of the simulation box (Table~\ref{tab:pbc}). Writing the system size in terms of the number of spheres that can fit end-to-end, $n_\sigma$, gives ${2R_I = n_\sigma \sigma}$, and thus 
\begin{equation}
N = \frac{\widehat{\varphi}dV\Gamma(\frac{d}{2} + 1)}{\pi^{d/2}}\bigg(\frac{n_\sigma}{2R_I}\bigg)^d.
\label{eq:systemSize}
\end{equation}
The limited role played by structural and dynamical correlations can separately be validated by verifying that the properties of the systems studied converge smoothly to those of $d\rightarrow\infty$ fluid, which are known exactly. 

In practice, we find that system sizes such that $n_\sigma \geq 2.2$ at $\widehat{\varphi}_\mathrm{d}$ might suffice. For $d\geq 15$, however, this criterion would require ${N \gg 10^5}$ 
which lies beyond the reach of our computational resources. 

\section{Dimensional Evolution of Fluid Structure}
\label{sec:dynamics}

In this section, we compare the dimensional evolution of structural observables with theoretical predictions. We specifically consider the fluid equation of state and the shell structure of $g(r)$.

\subsection{Equation of state}

\begin{figure*}[htp]
\includegraphics[width=0.9\linewidth]{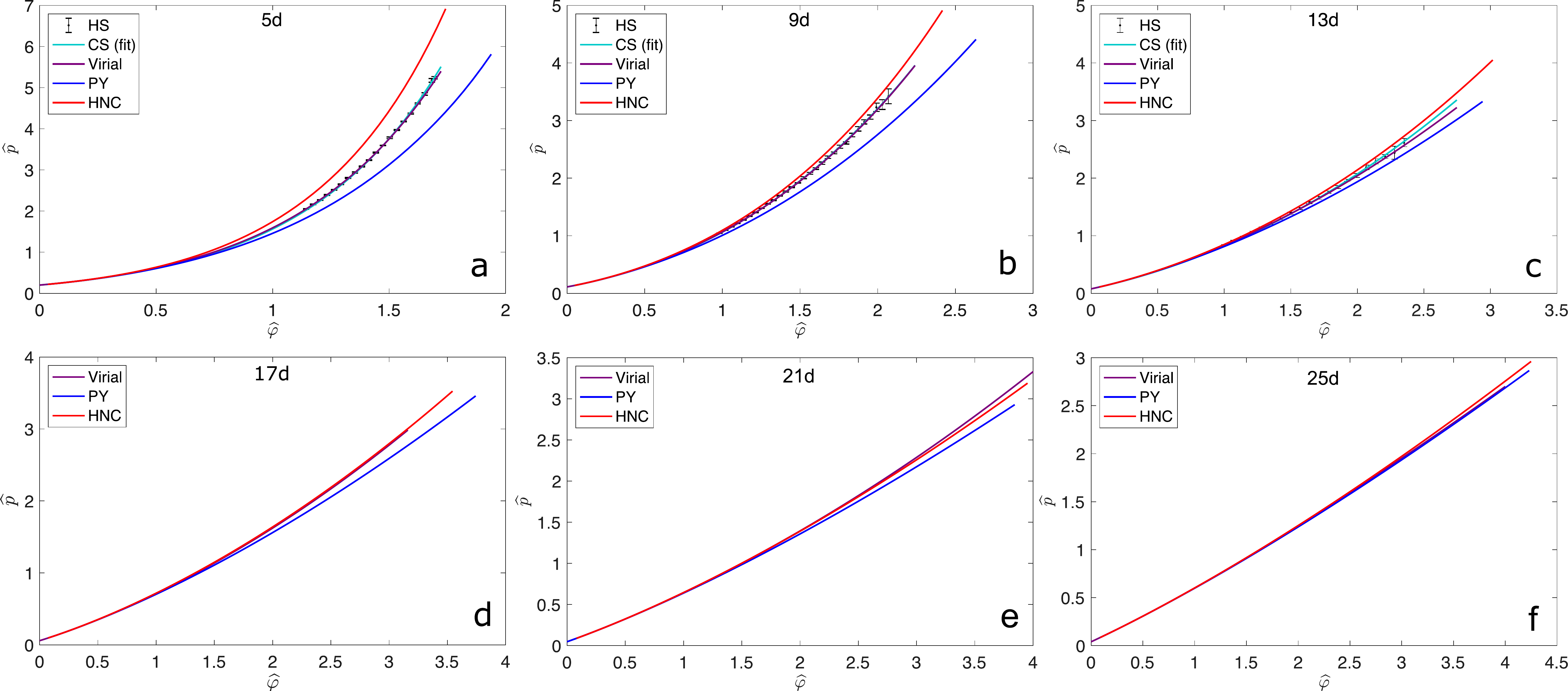}
\caption{Liquid equations of state for (a) $d=5$, (b) $d=9$, (c) $d=13$, (d) $d=17$, (e) $d=21$, (f) $d=25$. The first three panels compare simulation results (HS) with the fitted Carnahan-Starling form (CS), the Pad\'e resummed virial series, as well as the Percus-Yevick (PY) and Hypernetted Chain (HNC) predictions. Only the latter three quantities appear in the last three panels. From these results, it is clear that both the CS form and the resummed virial series remain good descriptors of the equation of state in that $\widehat{\varphi}$ regime in high $d$. It is also clear that the equation of state is systematically better described by HNC than by PY as $d$ increases. Remarkably, for $d\geq13$, HNC predictions have essentially converged with the simulation results.} \label{fig:CSCompare}
\end{figure*}

\begin{figure}[htp]
\includegraphics[width=0.85\linewidth]{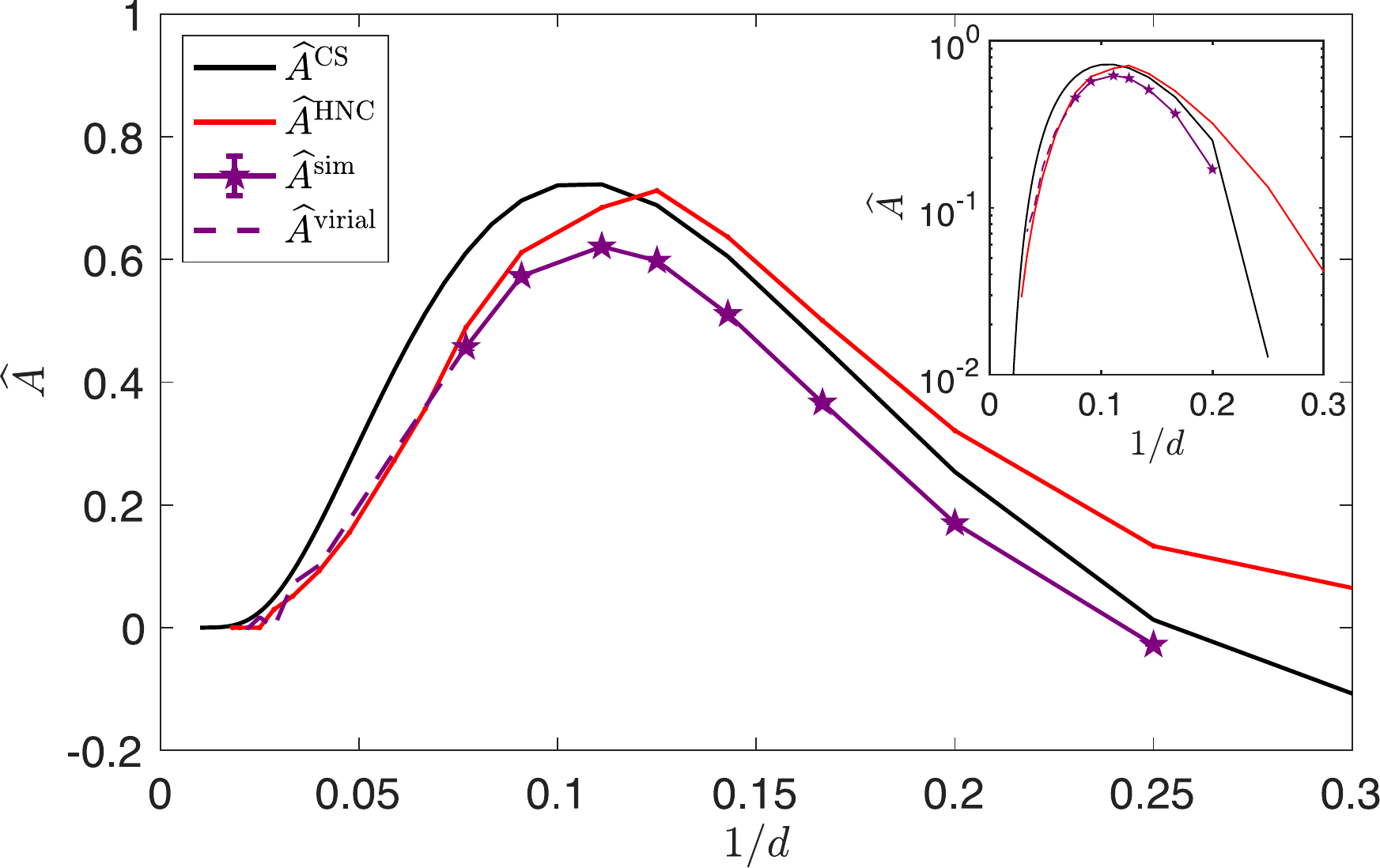}
\caption{Fitting coefficients for the CS form, $\widehat{A}^\text{sim}$ (purple stars), grow from a negative value for $d<5$ to a peak in $d=9$, before steadily decreasing. This behavior is consistent with prediction for $\widehat{A}^\text{CS}$~\cite{song_why_1989} (black), to which the numerical results could plausibly converge at even higher $d$ than is reached here. The quantity should therefore vanish as $d\rightarrow\infty$. The dimensional range is extended by fitting the virial (purple) and HNC (red)~\cite{zhang_computation_2014} up to $d=50$ to a CS form. High-$d$ convergence is best assessed on a logarithmic scale (inset).} 
\label{fig:AdFit}
\end{figure}

The liquid equation of state is one of the simplest observables to evaluate. Its relative crudeness is compensated by a dearth of available theoretical predictions. Here, we specifically compare numerical results (using Eq.~\ref{eq:eos}) with virial, PY, and HNC~\cite{ikeda_mode-coupling_2010} predictions. For further analysis, these results are succinctly fitted to a generalized Carnahan-Starling (CS) form~\cite{carnahan_equation_1969, song_why_1989, charbonneau_glass_2011},
\begin{equation}
\widehat{p} = \frac{1}{d} + \frac{\widehat{\varphi}}{2}\bigg[\frac{1+\widehat{A} \widehat{\varphi}}{(1-\varphi)^d}\bigg],
\label{eqn:csform}
\end{equation}
which has only one free parameter, $\widehat{A}$ (Fig.~\ref{fig:CSCompare}). This form captures numerical results quite well over the regime of interest, $\widehat{\varphi}_\mathrm{nf}<\widehat{\varphi}<\widehat{\varphi}_\mathrm{d}$.  (Higher-order corrections would mostly account for deviations at lower $\widehat{\varphi}$~\cite{ivanizki_generalization_2018}, and hence are not considered.) As expected, we find that $\widehat{A} \to 0$ as $d\to\infty$ thus recovering Eq.~\eqref{eqn:eosdinfty}. 
It is interesting to compare this fitted factor with the generalized CS treatment of Song et al.~\cite{song_why_1989}, for which 
\begin{equation}
\begin{split}
\widehat{A}^\text{CS} &= \frac{d}{2}\bigg(\frac{B_3}{B_2^2}\bigg)-\frac{d^2}{2^{d}}  = d\bigg[1- \frac{{}_2F_1\big(\frac{1}{2};\frac{1-d}{2};\frac{3}{2};\frac{1}{4}\big)}{\beta\big(\frac{1}{2},\frac{1+d}{2}\big)}\bigg] -\frac{d^2}{2^{d}}  \\&= dI_{\frac{3}{4}}(\frac{d+1}{2},\frac{1}{2}) - \frac{d^2}{2^{d}} \\& \sim \sqrt{\frac{6d}{\pi}}\bigg(\frac{3}{4}\bigg)^{d/2}\bigg[1-\frac{15}{4d} + \mathcal{O}\bigg(\frac{1}{d^2}\bigg)\bigg] - \frac{d^2}{2^{d}}
\end{split}
\end{equation}
where $B_n$ is the virial coefficient of order $n$, ${}_2F_1$ is the hypergeometric function, $\beta$ is the beta function, and $I_x(a,b)$ is the regularized incomplete beta function. A high-$d$ asymptotic scaling form is also provided. (Although this form could be developed to higher order for marginally better asymptotic scaling~\cite{nemes_uniform_2016}, the corrections would nevertheless become increasingly significant for $d<25$ because the $I_x(a,b)$ scaling only converges for $x\in[0,\frac{1}{2})$, even though an asymptotic expression can be obtained for $x\in[0,1)$.) This scaling form might give the impression that deviations (on a linear scale) from the $d\rightarrow\infty$ results appear markedly more significant for $d\lesssim50$, but convergence is actually smooth (see also Sec.~\ref{sec:phid}). Simulation estimates for $\widehat{A}$ from HS simulations (for $4\le d \le 11$), the virial (for $13 \le d \le 50$, see below), and HNC lend further support to this scaling form. The liquid-state approach to the $d\rightarrow\infty$ description is therefore highly non-trivial.  

We next consider the reliability of the virial series. For hard spheres coefficients have been reported up to $n=10$ in $4\le d \le 8$~\cite{clisby_ninth_2006}, and up to $n = 30$ for $d \ge 9$~\cite{zhang_computation_2014}. In order to make use of that series at high densities, however, it is necessary to resum it to capture its (apparent) divergence in the liquid phase. As is standard~\cite{sanchez_virial_1994, clisby_ninth_2006, bishop_equation_2008, lue_molecular_2021, charbonneau_thermodynamic_2021, charbonneau_three_2021}, we use $[\ell, m]$ Pad\'e approximants, which given the number of known terms in the virial series $n=\ell + m + 1$ are 
\begin{equation}
p = \frac{1 + \sum_{i=1}^\ell b_i\rho^i}{\sum_{i=1}^m \bar{b}_i\rho^i},
\label{eqn:pade}
\end{equation}
where $b_i$ and $\bar{b}_i$ are determined from the set of  coefficients $\{B_1,\dots, B_n\}$~\cite{guttmann_phase_1989}. There is no \textit{a priori} correct choice for $\ell$, $m$, and $n$; Pad\'e approximants are notoriously ill behaved, as evidenced by the emergence of spurious poles below $\widehat{\varphi}_\mathrm{d}$, for even small changes to $\ell$ and $m$. Nevertheless, conventional wisdom suggests the best fits occur for $\ell \approx m$~\cite{sanchez_virial_1994, clisby_ninth_2006, bishop_equation_2008, lue_molecular_2021, charbonneau_thermodynamic_2021, charbonneau_three_2021}. We thus here use $m-1 \le \ell \le m+1$ and choose the lowest order $\ell$ for which no spurious pole appears. Within these constraints, we find the $[4,5]$ approximant in $d \le 8$ and either the $[\ell,\ell]$ or the $[\ell,\ell-1]$ approximant  in $d > 8$ to most reliably converge. For $d\leq 13$ a solid agreement with numerical results is then obtained. 
It is thus reasonable to extract $\widehat{A}$ from the virial series up to $d=30$. For $d>30$, however, the relative error on the series coefficients makes this process numerically unreliable.

Having validated the numerical and resummed virial results, we can now consider the accuracy of the HNC and PY predictions. Both schemes properly converge to the $d\rightarrow\infty$ solution~\cite{wyler_hard-sphere_1987}, but their asymptotic scaling differs and it is unclear how well either describes high but finite-$d$ systems. We note that although HNC results are fairly crude in low dimensions, they recapitulate the numerical data increasingly well as $d$ increases. For $d=13$ the results are already pretty close, and for $d=17$ the virial and the HNC predictions overlap within the uncertainty on the resummation scheme (Fig.~\ref{fig:CSCompare}d). The expectation that HNC approaches the equation of state asymptotically well as $d$ increases~\cite{parisi_toy_2000, mangeat_quantitative_2016} is therefore here validated. By contrast, PY predictions remain quite far off the mark for $d<25$. They systematically undershoot both numerical and virial results. Hence, although PY does a remarkable (and likely fortuitous) job in $d=3$ and is correct in $d\rightarrow\infty$, its description of intermediate dimensions is inadequate. It cannot be relied upon for estimating other observables. 

The convergence of the HNC prediction onto the virial results as $d$ increases gives us confidence to consider the former more generally in higher $d$. In particular, a separate estimate of the fitting parameter $\widehat{A}$ can be obtained from HNC. Unlike the estimate from the virial series, this one is not affected by spurious poles, and hence its high-$d$ behavior is smoother (see Fig.~\ref{fig:AdFit}). The same overall trend is obtained. This combination of methods thus suggests that the asymptotic convergence to $\widehat{A}^{\mathrm{CS}}$ is rather slow (see Fig.~\ref{fig:AdFit} inset). 

\subsection{Pair distribution function}
\label{sec:structure}
\begin{figure*}[htp]
\includegraphics[width=0.9\linewidth]{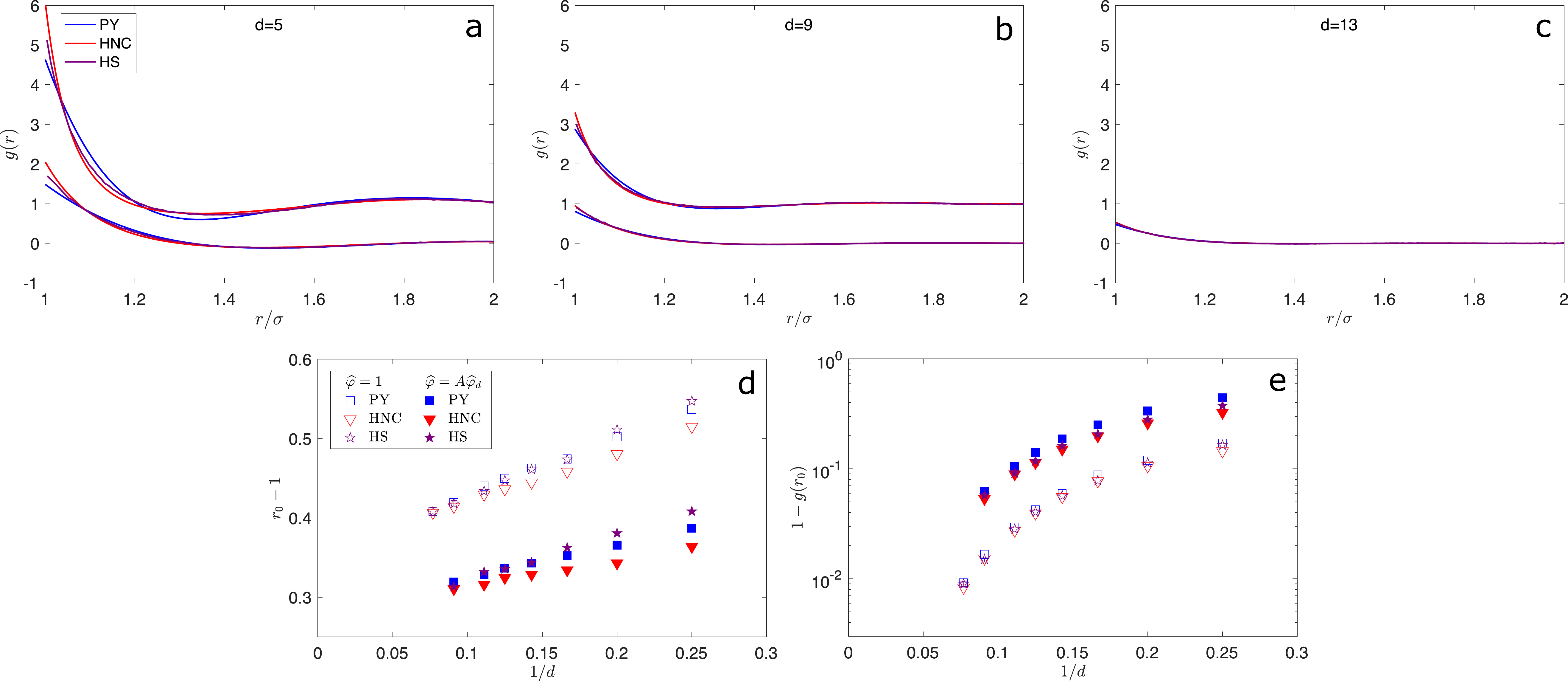}
\caption{Comparison of simulated $g(r)$ with PY and HNC predictions in (a) $d=5$, (b) $d=9$, and (c) $d=13$. Data are shown for both a fixed density near the onset, $\widehat{\varphi}=1$ (vertically offset by $-1$) and for a density near the simulation estimate for the dynamical transition, $A\widehat{\varphi}_\mathrm{d}$ with $A=0.93$. To quantify the visual differences between these descriptions we consider 
(d) the position $r_0$ and (e) depth $1-g(r_0)$ of the first minimum. In the $d\to\infty$ limit the depth and position of the first minimum of HS, PY, and HNC all seem to converge. Note that because systematic deviations appear in $d=11$ and $d=13$ at $\widehat{\varphi}=1$ for the system sizes considered in the rest of this work, they are here replaced with $N \rightarrow\infty$ values (see Appendix~\ref{sec:finSize}).}
\label{fig:gvsd}
\end{figure*}

Although the equation of state is directly related to contact features of the pair correlation (see Eq.~\eqref{eq:eos}), it provides no information about the liquid structure beyond that regime. In particular, it offers little insight into the first solvation shell, which includes the particles that control self-caging as density increases. Comparing HNC and PY with numerics in that range should therefore offer some insight about the quantitative reliability of $\widehat{\varphi}_\mathrm{d}$ predictions. 

Because $g(r)$ depends on density, a proper dimensional comparison should be scaled accordingly. Here, because we are interested in the co-evolution of structure and dynamics, we specifically monitor structure at two characteristic densities of sluggish liquids (further discussed in Sec.~\ref{sec:dynamics}): $\widehat{\varphi}_\mathrm{nf}\approx 1$ and $A\widehat{\varphi}_\mathrm{d}$. (The factor $A$ is chosen to be close to $\varphi_\mathrm{d}$ yet to remain computationally accessible for all $d$.) Figures~\ref{fig:gvsd}a-c are consistent with our expectation that $g(r)$ should deviate more strongly from a step function as density increase, but that these deviations should be systematically suppressed as $d$ increases. Both HNC and PY then also appear to capture numerical results increasingly well.

To quantify this convergence more carefully, we specifically consider the position, $r_1$, and depth, $g(r_1)$, of the first minimum of $g(r)$, which traditionally delimits the end of the first solvation shell. 
Because $g(r)$ tends toward a step function as $d\rightarrow\infty$, $1-g(r_1)$ should then vanish, however $r_1-1$ is not similarly constrained.  
Figure~\ref{fig:gvsd} shows that as $d\to \infty$ the former quantity indeed vanishes--and does so monotonically--and that the latter doesn't. The thickness of the first solvation shell, which in low $d$ commonly describes the caging process, therefore does not asymptotically estimate the cage size, which vanishes instead as $1/d$~\cite{parisi_mean-field_2010}. 

Interestingly, although predictions for both quantities from PY and HNC converge as $d\rightarrow \infty$, 
numerical simulation results generally appear to converge toward HNC predictions more directly as $d$ increases. The effect is marked and systematic for the minimum depth. The minimum position drifts from being above the PY prediction to near the HNC one as $d$ increases, which here as well suggests that the low-$d$ agreement with PY might be due to a fortuitous cancellation of error. This analysis thus further supports the advantageous asymptotic dimensional scaling of HNC in describing structural pair correlations in the liquid.
Given that computationally accessible system sizes for $d=13$ at densities beyond onset are markedly too small to capture the shell structure (see Appendix~\ref{sec:finSize}), however, we do not consider these systems further.


\section{Dimensional evolution of fluid dynamics}
\label{sec:dyn}
In this section, we compare the dimensional evolution of dynamical observables with theoretical predictions. Specifically, we consider the dynamical onset and $\widehat{\varphi}_\mathrm{d}$.

\subsection{Dynamical onset}
In order to correlate these structural observations with liquid dynamics, we first characterize the dynamical onset. To define this onset, recall that at low densities, hard spheres that evolve under a MC (Brownian-like) dynamics are Fickian fluids with a MSD given by $\langle r^2(t) \rangle = Dt$, where $D$ is a constant set by microscopic dynamics alone. In other words, the MSD is then featureless.
As density increases, however, the fluid becomes non-Fickian, thus hinting at the emergence of non-trivial dynamics. This transformation, however, is but a crossover even in mean-field descriptions, and hence it is not uniquely defined. We here follow the convention of Refs.~\cite{charbonneau_hopping_2014,manacorda_numerical_2020} in identifying the dynamical onset from the emergence of an inflection point in the (logarithmically scaled) MSD 
\begin{equation}
\frac{\mathrm{d}^2(\ln\langle r^2(t) \rangle)}{\mathrm{d}(\ln t)^2} = 0.
\end{equation}

The resulting estimate is consistent with $\widehat{\varphi}_\mathrm{nf} =1.3(1)$ in all $d$, which agrees remarkably well with molecular dynamics results~\cite{charbonneau_hopping_2014} (see Table~\ref{table:phiTable}). The choice of microscopic dynamics therefore appears to have but a relatively weak effect on that onset, at least in the $d$ range studied. By contrast, recent work by Manacorda et al.~finds an onset slightly below unity for both Newtonian or Brownian dynamics for $d\rightarrow\infty$~\cite{manacorda_numerical_2020},  in agreement with numerical results for minimally structured yet finite-$d$ (Mari-Kurchan) models. It therefore seems natural to conclude that the structure of HS fluids underlies the $\sim30\%$ difference between finite-$d$ and $d\rightarrow\infty$ results. But what particular structural feature could explain this difference? The shell structure of low-$d$ HS fluids may be responsible, but one would then (naively) expect the dynamical onset to emerge at a \emph{lower} $\widehat{\varphi}$ than what the $d\rightarrow\infty$ solution predicts. The difference between finite-$d$ and $d\rightarrow\infty$ results is nevertheless rather small, especially relative to the marked increase of $\widehat{\varphi}_\mathrm{d}$ with $d$ (see Sec.~\ref{sec:phid}). Both the structural origin and the meek dimensional dependence of $\varphi_\mathrm{nf}$ are therefore puzzling. Without a crisper understanding of the finite-$d$ mean-field dynamics, further insight likely remain out of reach. 

\begin{figure}[h]
\includegraphics[width=0.85\linewidth]{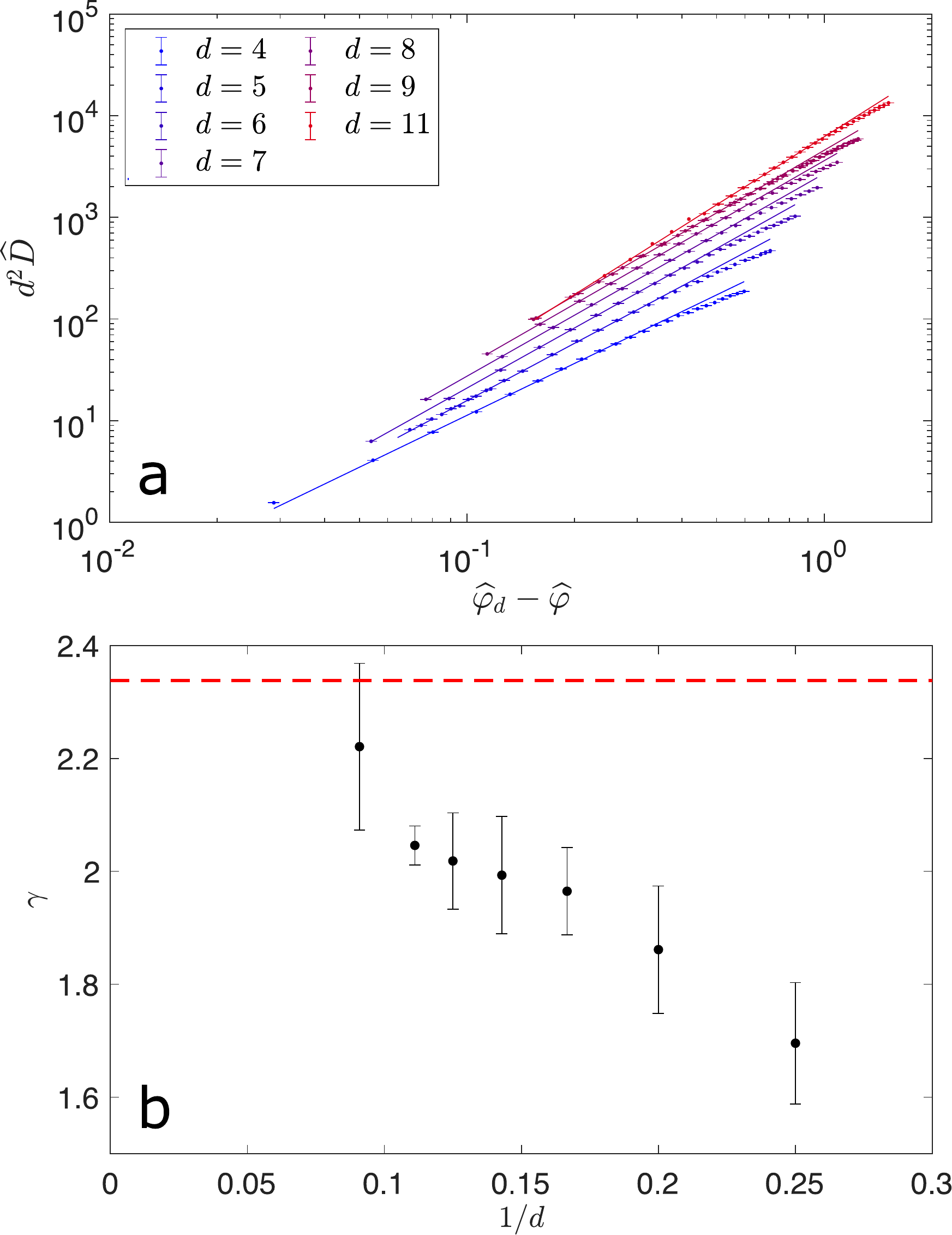}
\caption{The (pseudo-)critical scaling of the scaled diffusivity $\widehat{D}$ from Eq.~\eqref{eqn:tauAlpha} (offset vertically by a factor of $d^2$ for clarity) over the relevant density regime (see text) provides estimates for both \textbf{a)} the dynamical transition density $\varphi_\mathrm{d}$ and \textbf{b)} the associated critical exponent $\gamma$. For $d\geq4$, $\gamma$ rises monotonically and appears to saturate at the $d\rightarrow\infty$ value of $\gamma=2.33786$~\cite{kurchan_exact_2013} (red dashed line). Numerical values of $\widehat{\varphi}_\mathrm{d}$ and $\gamma$ are reported in Table~\ref{table:phiTable}.}
\label{fig:tauAlpha}
\end{figure}

\subsection{Dynamical transition}
\label{sec:phid}
We finally consider the dynamical transition regime. At a genuine dynamical transition, perfect caging would result in an infinitely-extended MSD plateau. In addition, upon approaching the transition density the diffusivity and the structural relaxation time would diverge, scaling critically as
\begin{equation}
D^{-1}\sim\tau_\alpha \sim (\varphi_\mathrm{d}-\varphi)^{-\gamma}.
\label{eqn:tauAlpha}
\end{equation}
%
Like any spinodal critical point, however, a true dynamical transition can only exist in the mean-field limit of high-spatial dimension, $d\to \infty$. In finite-$d$ systems, activated processes blur the singularity into a crossover, and decouple $D$ and $\tau_\alpha$ thus giving rise to a breakdown of the Stokes-Einstein relation~\cite{charbonneau_hopping_2014,adhikari_spatial_2021}. If activated processes can be screened somehow, or if dimension is sufficiently high, however, traces of criticality may still be distinguished, including the pseudo-critical scaling of Eq.~\eqref{eqn:tauAlpha}. Because $\tau_\alpha$ naturally screens some of the activated processes~\cite{bouchaud_aging_1997, bouchaud_mode-coupling_1996, schweizer_entropic_2003, mirigian_elastically_2014}, its scaling is commonly used to explore this regime (Fig.~\ref{fig:tauAlpha}a). However, the value of $\tau_\alpha$ depends on both $a$ and the functional form of the overlap function in Eq.~\eqref{eqn:overlapFunction}, which themselves should evolve with $d$. In this context, we prefer to consider $D$ for comparing dynamics across dimensions. 
It is here extracted from fitting the long-time scaling of the MSD 
using an empirical decay form  
\begin{equation}
\langle r^2(t) \rangle/t = D + a_0 t^{-a_1},
\end{equation}
where $a_0$ and $a_1$ are positive numbers that depend on $\varphi$, and are unimportant for the subsequent analysis. 

This approach is nevertheless not completely devoid of difficulties.  For MC dynamics, in particular, at microscopic times $t \lesssim \mathcal{O}(1)$ the system behavior is roughly independent of density. At asymptotically small $D$, this effect is negligible, but given the limited available dynamical regime this effect cannot be brushed aside. In order to calibrate our results, we thus use of a multiplicative scale factor that fixes the MSD at $t=1$ for all densities considered. In addition, the pseudo-critical scaling regime of Eq.~\ref{eqn:tauAlpha} can only be expected to hold between $\varphi_\mathrm{nf}$ and the density at which the Stokes-Einstein relationship (SER) is found to break $\varphi_\mathrm{SER}$ (see Ref.~\cite{charbonneau_hopping_2014}). 
Even this range is too broad. The $d\rightarrow\infty$ scaling of $D$ deviates markedly from the asymptotic power-law scaling already at $\widehat{\varphi} \sim \widehat{\varphi}_\mathrm{d}/2$~\cite{manacorda_numerical_2020}.
Using this observation as a guide, we here only fit the upper half of the density regime beyond the onset of non-Fickian diffusion to the (pseudo-)critical scaling form of Eq.~\eqref{eqn:tauAlpha}, while staying clear of the SER breakdown regime in low $d$ (see Fig.~\ref{fig:tauAlpha}). (In practice, we use $\widehat{\varphi}>(1 + \widehat{\varphi}_\mathrm{d})/2$.) 

\begin{table}
\begin{tabular}{ c | c | c | c | c }
\hline
\hline
$d$ & $\widehat{\varphi}_\mathrm{d}$ & $\widehat{\varphi}_\mathrm{nf}$ & $\gamma$ & Ref.\\
\hline
4 & 1.624 & -- & 2.4(3)  & ~\cite{charbonneau_numerical_2010}\\
& 1.60(2) & -- & -- & ~\cite{charbonneau_glass_2011}\\
& 1.6144(8) & 1.17(2) & 1.92(3) & ~\cite{charbonneau_hopping_2014}\\
& \textbf{1.597(7)} & \textbf{1.30(2)} & \textbf{1.70(11)} & \\
\hline
5 & 1.71(3) & -- & -- & ~\cite{charbonneau_glass_2011}\\
& 1.7171(6) & 1.22(13) & 1.95(3) & ~\cite{charbonneau_hopping_2014}\\
& \textbf{1.706(7)} & \textbf{1.36(8)} & \textbf{1.87(10)} & \\
\hline
6 & 1.83(5) & -- & -- & ~\cite{charbonneau_glass_2011}\\
& 1.8379(11) & 1.17(11) & 2.00(3) & ~\cite{charbonneau_hopping_2014}\\
& \textbf{1.829(5)} & \textbf{1.33(5)} & \textbf{1.96(8)} & \\
\hline
7 & 1.94(9) & -- & -- & ~\cite{charbonneau_glass_2011}\\
& 1.968(2) & 1.19(9) & 2.0(1) & ~\cite{charbonneau_hopping_2014}\\
& \textbf{1.956(9)} & \textbf{1.29(10)} & \textbf{1.99(10)} & \\\hline
8 & 2.07(2) & -- & -- & ~\cite{charbonneau_glass_2011}\\
& 2.107(2) & 1.28(6) & 2.15(5) & ~\cite{charbonneau_hopping_2014}\\
& \textbf{2.087(10)} & \textbf{1.36(8)} & \textbf{2.02(9)} & \\\hline
9 & 2.19(3) & -- & -- & ~\cite{charbonneau_glass_2011}\\
& \textbf{2.225(5)} & \textbf{1.28(3)} & \textbf{2.05(3)} & \\\hline
10 & 2.31(5) & -- & -- & ~\cite{charbonneau_glass_2011} \\\hline
11 & 2.44(9) & -- & -- & ~\cite{charbonneau_glass_2011}\\
& \textbf{2.53(3)} & \textbf{1.18(4)} & \textbf{2.22(15)} & \\\hline
12 & 2.6(1) & -- & -- & ~\cite{charbonneau_glass_2011} \\\hline
\end{tabular}
\caption{Pseudo-critical parameters for the dynamical transition as well as the dynamical onset density from this work (bold), along with estimates from Refs.~\cite{charbonneau_numerical_2010,charbonneau_glass_2011, charbonneau_hopping_2014}.}
\label{table:phiTable}
\end{table}

The resulting estimates of $\widehat{\varphi}_\mathrm{d}$, and of the non-universal critical exponent $\gamma$ can be seen in Table \ref{table:phiTable} and Fig.~\ref{fig:tauAlpha}. 
The resulting $\widehat{\varphi}_\mathrm{d}$ are marginally different, but generally consistent. They also follow the same trend as previously reported values~\cite{charbonneau_glass_2011,charbonneau_hopping_2014}. The results for $\gamma$ obtained via Monte Carlo dynamics also agree, within error bars, with the molecular dynamics estimates~\cite{charbonneau_glass_2011,charbonneau_hopping_2014}, and extend by nearly 50\% the dimensional range of under solid numerical control. 

The larger error bars reported here reflect the increased understanding of systematic errors involved in simulating glass forming liquids and extracting the pseudo-critical behavior. We thus expect the current estimates to hold for the foreseeable future. The simultaneous fit of $\widehat{\varphi}_\mathrm{d}$ and $\gamma$, however, leaves some undue wiggle room in the analysis. Because the resulting power-law scaling depends fairly sensitively on $\widehat{\varphi}_\mathrm{d}$, which itself increases fairly rapidly with $d$, finite-$d$ theoretical estimates for the quantity are unlikely to ever serve as efficient substitutes. By contrast, $\gamma$ increases slowly and monotonously with $d$ towards its $d\rightarrow\infty$ value, $\gamma=2.33786$~\cite{kurchan_exact_2013}. If a finite-$d$ estimate of that exponent were independently determined (using, e.g., the approach of Ref.~\cite{caltagirone_dynamical_2012}), then a more controlled test of the theory and a more reproducible estimate of $\widehat{\varphi}_\mathrm{d}$ would be possible. 

\begin{figure}[h]
\includegraphics[width=0.85\linewidth]{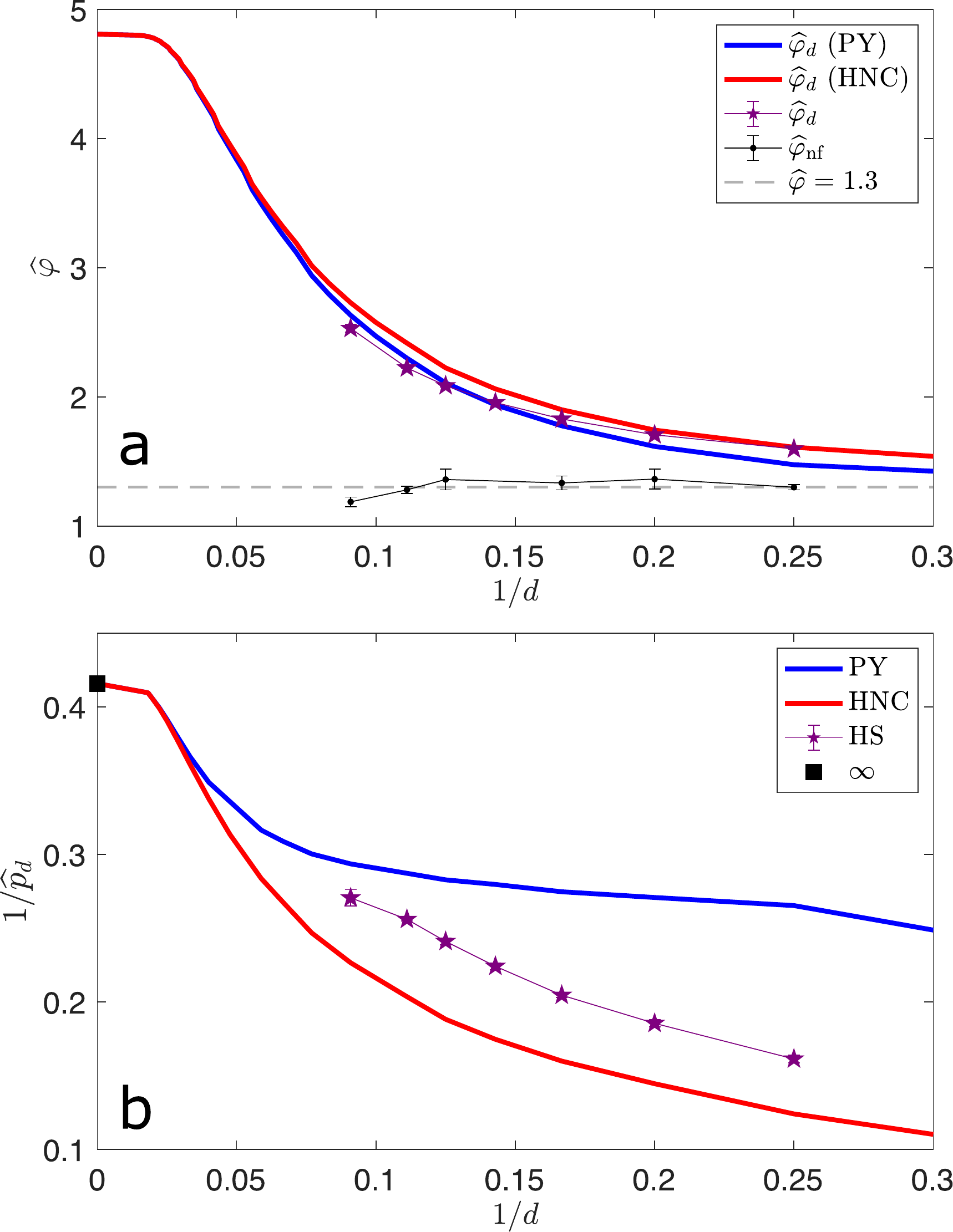}
\caption{Dimensional evolution of \textbf{a)} the dynamical transition $\widehat{\varphi}_\mathrm{d}$ and dynamical onset $\widehat{\varphi}_\mathrm{nf}$ as well as \textbf{b)} $\widehat{p}_\mathrm{d}$. For the dynamical transition values, the predicted mean-field values for both PY and HNC are given, while the onset should here be compared with the mean-field prediction $\widehat{\varphi}_\mathrm{nf} \le 1$. The dynamical transition systematically deviates from the mean-field estimate obtained using PY and HNC~\cite{mangeat_quantitative_2016}, but generally follows the HNC trend more reliably. 
}
\label{fig:phiDphiOnset}
\end{figure}

Mangeat and Zamponi~\cite{mangeat_quantitative_2016} have shown that both HNC and PY provide correct order-of-magnitude estimates for $\widehat{\varphi}_\mathrm{d}$ in low $d$, and converge with each other for $d \gtrsim 30$. Simulation results, however, show that while both HNC and PY capture the general trend of $\widehat{\varphi}_\mathrm{d}$ in finite $d$, both overshoot its values beyond even our enlarged error bars (Figure~\ref{fig:phiDphiOnset}a) in the highest $d$ attained. One potential source of discrepancy is the reliance of Mangeat and Zamponi on the Gaussian cage approximation, which is violated in finite-$d$ hard spheres~\cite{charbonneau_hopping_2014, adhikari_spatial_2021} and is known to lead to significant $1/d$ corrections in a related model~\cite{biroli_interplay_2021,biroli_JCP_2021,biroli_local_2021}. The dimensional range accessible in simulations is however too small to assess this hypothesis. Structural deviations at lower $d$ are indeed too pronounced for their dynamical contribution to be ascertained.

\begin{figure}[h]
\includegraphics[width=0.85\linewidth]{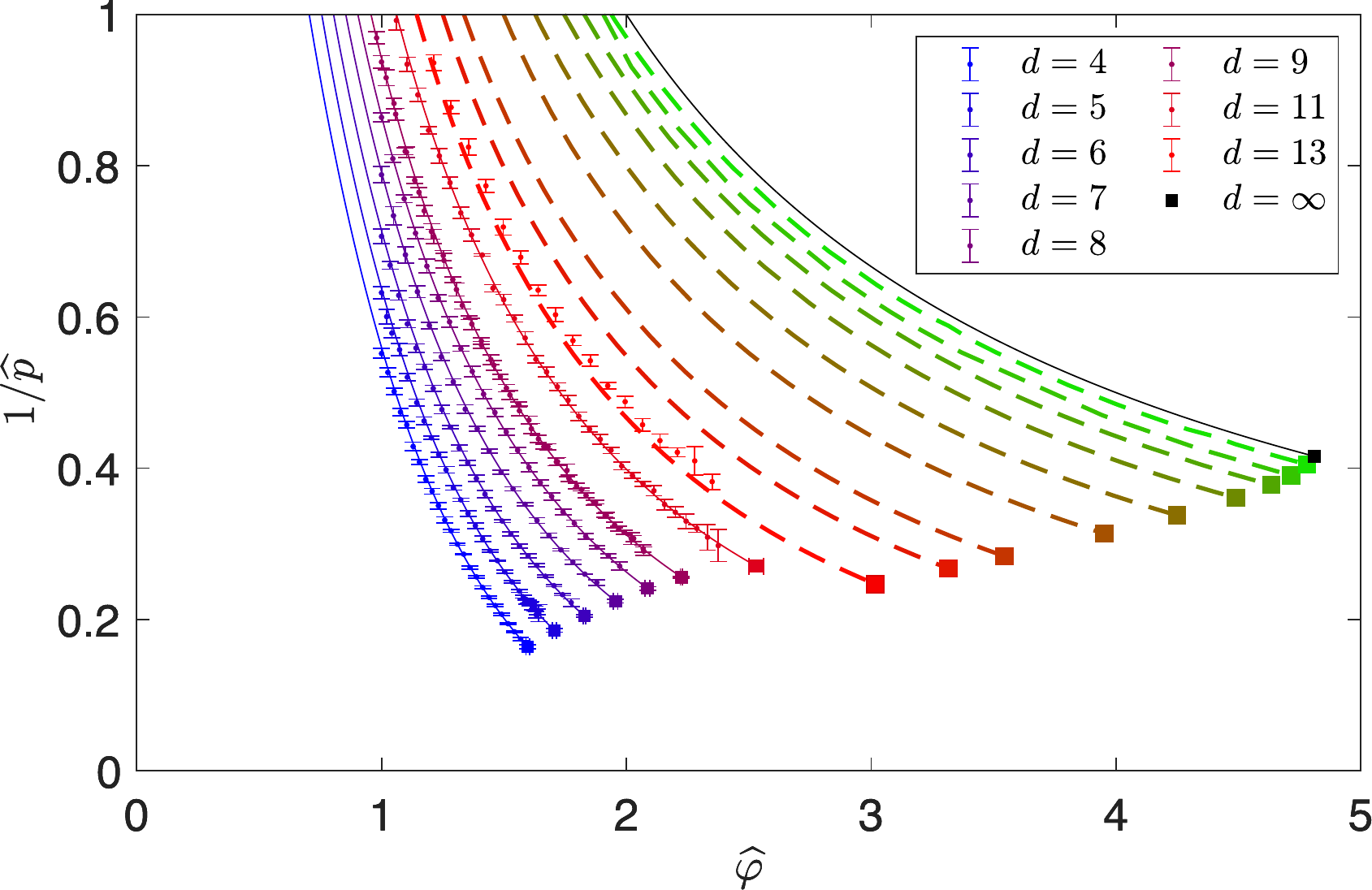}
\caption{Liquid equations of state for $d=4$-$9$, $11$, and $13$ (graded from blue to red) obtained by fitting simulation results (symbols) to the generalized CS form in Eq.~\eqref{eqn:csform}, along with the $d\rightarrow\infty$ equation of state from Eq.~\eqref{eqn:eosdinfty} (black line). Squares denote the dynamical transition obtained as in Sec.~\ref{sec:dynamics}, and $\widehat{\varphi}= 4.81\ldots$ for $d\rightarrow\infty$~\cite{parisi_mean-field_2010}. HNC equations of state for $d=15$, $17$, $21$, $25$, $30$, $35$, $40$, and $50$ (dashed lines graded from red to green) are depicted up to the predicted dynamical transition. In $d=13$, simulation pressure data is shown, but the CS fit is omitted, and the HNC equation of state is shown instead (dashed red line).}
\label{fig:invertedEOS}
\end{figure}

In any event, considering the success of HNC in predicting the equation of state and the pair structure, our results are consistent with the suspicion of Mangeat and Zamponi that HNC estimates for $\widehat{\varphi}_\mathrm{d}$ (and the corresponding pressure $\widehat{p}_\mathrm{d}$) are \textit{at best} an upper bound. Similarly, the inadequacy of PY in capturing the equation of state and the pair structure at intermediate $d$ results in predictions for $\widehat{\varphi}_\mathrm{d}$ that overshoot while $\widehat{p}_\mathrm{d}$ undershoots the numerical results. 



%

\section{Conclusion}
\label{sec:conclusion}

%

Our results suggest that the gap in structure between the $d\rightarrow\infty$ predictions and numerical simulations is essentially closed. As dimension increases, HNC tracks the smoothing of the pair correlation function and nearly quantitatively captures simulation and virial results for $d\gtrsim13$. 
(PY, however, fails to capture key structural features in intermediate dimensions, $d=4$-$13$.) 
Given this match, we also understand that non-trivial structural corrections are almost irrelevant for $d\gtrsim 50$, and that higher-order correlations are necessary to quantitatively describe systems with $d<13$. 

Our results, however, also suggest that a certain gap in the dynamical description persists (see Fig.~\ref{fig:invertedEOS}). Mean-field theory-based predictions for the dynamical onset and the dynamical transition are close to the simulation results, but remain quantitatively distinct (even when the pair structure description is nearly flawless). Can more be done to bridge the disconnect? Unfortunately, numerical techniques seem to have approached their practical limit. Properly simulating higher $d$ fluids would require either an enormous computational undertaking or a marked shift in numerical techniques. In the near term, theoretical improvements are more likely. In particular, a first-principle calculation for $\gamma$ should  be within reach, and advances on the dynamical mean-field theory~\cite{liu_dynamics_2021} might eventually provide some quantitative insight. Subtle higher-order correlations could also impact the mean-field-like dynamics, as has recently been suggested~\cite{ridout_correlation_2020} is also possible, but cannot be evaluated without a more robust dynamical theory.





\begin{figure*}[htb]
\includegraphics[width=0.8\linewidth]{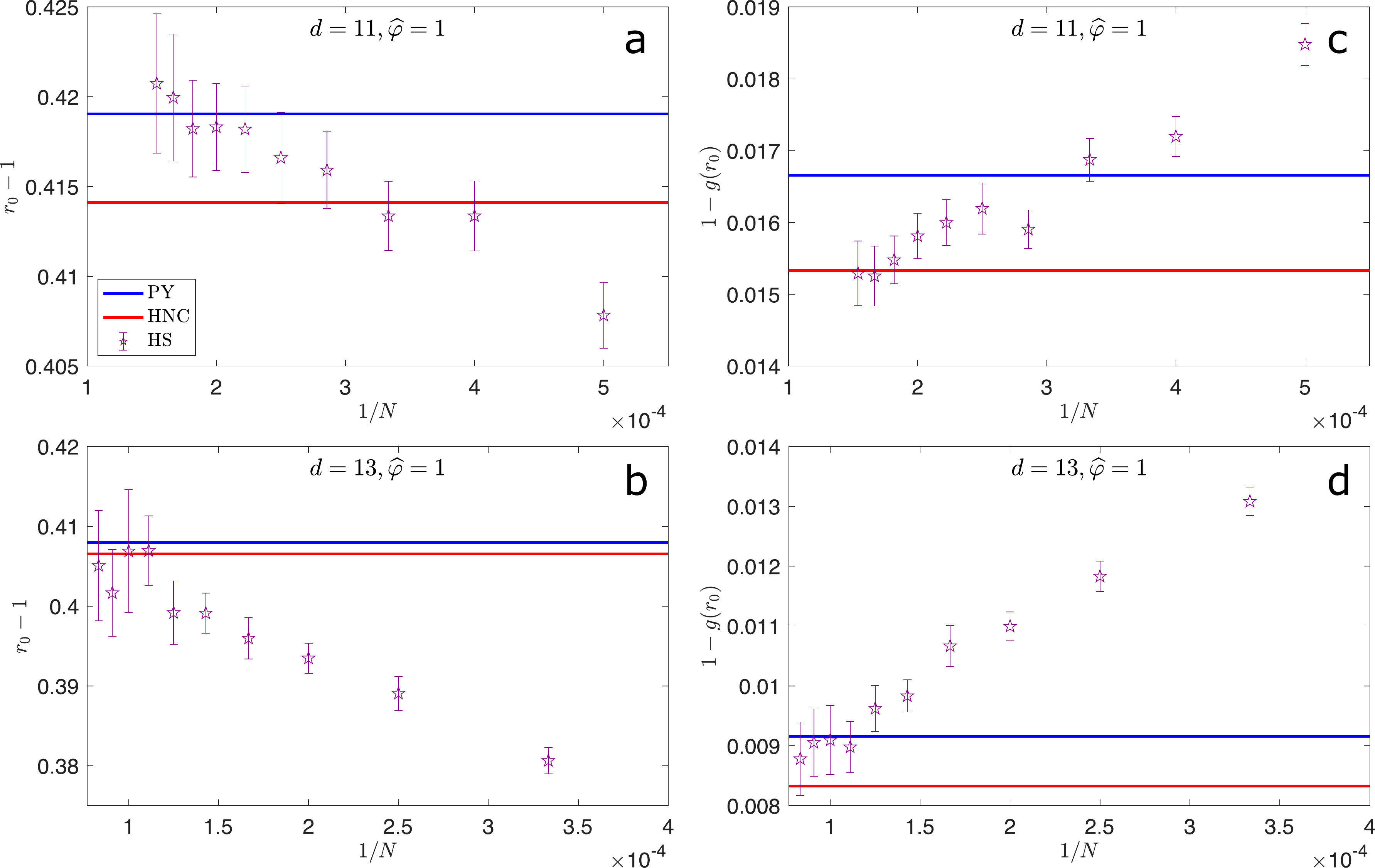}
\caption{Finite-size scaling of the position of the first minimum of the pair correlation function $r_0 - 1$ in \textbf{a)} $d=11$ and \textbf{b)} $d=13$ and the value that the pair correlation takes in \textbf{c)} $d=11$ and \textbf{d)} $d=13$, all taken at $\widehat{\varphi} = 1$. In both dimensions, the system size used in this work is the second smallest reported result. The results quickly converge as $N$ increases. The discrepancy is deemed small in $d=11$, but significant in $d=13$. PY and HNC values are reported to facilitate the comparison with Fig.~\ref{fig:gvsd}.}
\label{fig:gofrFS}
\end{figure*}

\begin{acknowledgements}

We thank Nathan Clisby for useful discussions about the virial expansion and Francesco Zamponi for various discussions. We would also like to thank Atushi Ikeda for sharing a code that solves the HNC structure and Andres Santos for a code that solve the PY structure. This work was supported by grants from the Simons Foundation (Grant No.~454937 to P.C.) The simulations were performed at both Duke Compute Cluster (DCC)---for which the authors thank Tom Milledge’s assistance---and on Extreme Science and Engineering Discovery Environment (XSEDE), which is supported by National Science Foundation grant number ACI-1548562. Data relevant to this work have been archived and can be accessed at the Duke Digital Repository~\cite{data}.
\end{acknowledgements}


%
%

\appendix

\section{Finite-size scaling in $d=11$ and $13$}
\label{sec:finSize}

The structure and dynamics of supercooled liquids generally depend only weakly on system size, provided that the simulation box is large enough to contain a representative environment for a given particle. However, in high dimensions, Eq.~\ref{eq:systemSize} shows that the system size necessary to overcome these effects grows exponentially with  $d$. Simulation results are thus most sensitive to system size in the highest $d$ considered, $d=11$ and $13$, where the limits of current computational feasibility are approached.

In Fig.~\ref{fig:gofrFS}, we show how the first minimum of $g(r)$ at the onset density evolves with $N$ (as in Fig.~\ref{fig:gvsd}). In both cases, the system size used in this work is the second smallest reported. Upon increasing system size, the structural observables appear to plateau at larger $N$ than what is used here, indicating that significant corrections to the structure (and dynamics) persist. For $d=13$ this discrepancy is quite notable, but for $d=11$, it is relatively small. Only the latter is thus included in the subsequent dynamical analysis. 

\bibliography{highDStruct,footnotes}

\end{document}